\documentclass[prd,reprint,showpacs,showkeys]{revtex4-1}
\pdfoutput=1
\usepackage{amsfonts}
\usepackage{mdframed}
\usepackage{amssymb}
\usepackage{amsmath}
\usepackage{graphicx}
\usepackage[font={footnotesize,it}]{caption}

\begin{document}

\title{Emergent cosmological constant from colliding electromagnetic waves}
\author{M. Halilsoy}
\email{mustafa.halilsoy@emu.edu.tr}
\author{S. Habib Mazharimousavi}
\email{habib.mazhari@emu.edu.tr}
\author{O. Gurtug}
\email{ozay.gurtug@emu.edu.tr}
\date{\today }
\affiliation{Department of Physics, Eastern Mediterranean University, Gazima\v{g}usa,
north Cyprus, Mersin 10 - Turkey}

\begin{abstract}
In this study we advocate the view that the cosmological constant is of
electromagnetic (em) origin, which can be generated from the collision of em
shock waves coupled with gravitational shock waves. The wave profiles that
participate in the collision have different amplitudes. It is shown that,
circular polarization with equal amplitude waves does not generate
cosmological constant. We also prove that the generation of the cosmological
constant is related to the linear polarization. The addition of cross
polarization generates no cosmological constant. Depending on the value of
the wave amplitudes, the generated cosmological constant can be positive or
negative. We show additionally that, the collision of nonlinear em waves in
a particular class of Born-Infeld theory also yields a cosmological constant.
\end{abstract}

\maketitle

\section{Introduction}

The subject of colliding plane waves (CPW) in general relativity constitutes
one of the important topics that the effects of the nonlinearity of the
Einstein's equations manifests itself explicitly. The basic results of CPW's
are not limited to find exact solutions, but rather its connections with
other predictions of the theory of general relativity such as the spacetime
singularities and the black hole interiors. (see \cite{1}, for a general
review of related works). Although the collision of plane waves assumes
idealized situations (the waves that participate in the collision are
assumed to be plane symmetric, having an infinite extent in transverse
directions), the dynamic nature of CPW spacetime may provide a theoretical
background to experimental observations.

For example, according to the Standard Big Bang Cosmological Model in which
the universe contains a cosmological constant, the universe went through an
exponential growth called inflation and caused the formation of ripples
propagating at the speed of light in the fabric of spacetime, called the
gravitational waves within a tiny fraction of time after the big bang. It is
now well understood that electromagnetic (em) radiation decoupled from free
electrons about 380,000 years after the big bang \cite{2}. Once they formed,
as a requirement of the Einstein's theory of relativity, em waves are
coupled with primordial gravitational waves and naturally their nonlinear
interactions started to shape the em distribution.

The findings of BICEP-2, can be given as an example to such phenomena \cite%
{BICEP}. The observed B-mode in the polarization vector of the cosmic
microwave background (CMB) radiation may be explained as a result of
interaction with the primordial gravitational waves originated during the
inflationary phase of the universe. This problem can be considered within
the context of CPW and the exact analytic solution to the Einstein-Maxwell
equations. The nonlinear interaction between plane gravitational waves and
shock em waves with cross polarization is of utmost importance. Here, the
primordial gravitational waves are assumed to be impulsive and shock types
for the sake of an analytic exact solution. It has been shown in \cite{3},
that the Faraday rotation in the polarization vector of em waves can be
attributed to the encounters with the strong gravitational waves with cross
polarization.

On the other hand, the cosmological constant in the Standard Model of Big
Bang Cosmology has been associated with the dark energy. Understanding the
origin of the cosmological constant, its role in the universal vacuum
energy, its repulsive effect in the accelerating expansion of the universe
and related matters all constitute a vast literature in modern cosmology.
Although experimental observations revealed much information about the
evolution of the universe, the origin of the cosmological constant still
lacks a satisfactory answer.

Recently, within the framework of CPW, one possible mechanism about the
origin of the cosmological constant has been introduced by Barrabes and
Hogan \cite{4}. In this study, it has been shown that, the cosmological
constant emerges as a result of nonlinear interaction of plane
electromagnetic (em) shock waves accompanied by gravitational shock waves.
As it was given in \cite{4}, this is a \textit{special solution} in the
sense that, there is only one component of electric and magnetic fields of
the combined em waves that participate in the collision. On the other hand,
the fundamental solution in this context is the Bell-Szekeres (BS) \cite{5}
solution which describes the collision of plane em shock waves in which
there are two components of equal amplitudes of electric and magnetic fields
that participate in the collision. Thus, the solution given in \cite{4}, is
a special case that has no BS limit of equal em amplitudes \cite{5}.

Our motivation in this paper is to explore in detail the effects of
polarization and wave amplitudes on the emergence of the cosmological
constant as a result of nonlinear interaction of plane em shock waves
coupled with gravitational shock waves. That is, our strategy from the
outset, is not to introduce a cosmological constant in the initial data of
the problem but rather to obtain it emergent as a result of colliding em
data. Owing to the importance of the problem, we wish to extend the solution
presented in \cite{4} further in various directions. First, we consider the
nonlinear interaction of em waves with different amplitudes in the initial
data of BS solution. We show that emergent cosmological constant relates
only to the linear polarization context of the waves with different
amplitudes. We obtain that the cosmological constant $(=\lambda _{0})=\alpha
^{2}-\beta ^{2}$, where $\alpha $ and $\beta $ denote the amplitude
constants of electric field components along $x$ and $y$ directions,
respectively. Such a theoretical prediction probably may be verified by
experimental observations and this naturally necessitates to reevaluate the
polarization data of CMB.

Extension of the BS solution to cross-polarized collision with single
essential parameter was also found \cite{6,7}. In this problem the two
incoming waves have non-aligned polarization vectors prior to the collision
and naturally give rise to an off-diagonal component in the metric. This is
analogous to the relation of Khan-Penrose \cite{8} and Nutku-Halil \cite{9}
metrics. Secondly, in order to understand the effect of polarization
together with different amplitudes, we extend the linear polarization
problem to the case of cross-polarization, however, this doesn't yield a
pure cosmological constant term. Instead, we obtain a general
energy-momentum without immediate interpretation but yet it can be
considered as a conversion of em energy into other forms.

In addition, we answer the question whether colliding em waves accompanied
with gravitational waves in gravity coupled nonlinear electrodynamics \cite%
{10} give rise to cosmological constant or not. Our finding for the
Born-Infeld (BI) \cite{11} theory is positive, however, this leaves the case
of different nonlinear electromagnetic models open.

The work described in this paper and also in \cite{4}, i.e. generation of
cosmological constant in the interaction region can be explained as a result
of the re-distribution of the incoming energies in the waves that
participate in the collision. Furthermore, as a by product besides the
cosmological constant, two light-like shells are also generated on the null
boundaries accompanying the impulsive gravitational waves. To recall a
similar scenario and seek support from a different (i.e. quantum) domain of
physics we refer to the theoretical side to the historic Breit-Wheeler
analysis \cite{12} of matter creation from the process of photon collisions.
On the experimental side, this has been taken seriously in recent times
through energetic laser photon collisions to materialize the idea at a grand
scale \cite{13}. If this quantum picture has any reflections in our
macroworld it must correspond with our approach of colliding em plane waves,
which is entirely classical.

Let us add that in addition to Ref. \cite{4}, Barabes and Hogan also gave a
method to generate a cosmological constant from collision of pure
gravitational shock waves \cite{14}. In this work the energy-momentum
created with the cosmological constant balances with the emergent null
currents on the boundaries of the collision. Hence, the consistency of the
Einstein's equations hold.

The organization of the paper is as follows. Section II explores the
collision of shock waves in Einstein-Maxwell (EM) theory. Section III
considers collision of waves in nonlinear electrodynamic. The paper ends
with Conclusion in Section IV. In Appendix A / B, we provide all Ricci /
curvature components. Appendix C shows the effect of cross polarization
while Appendix D presents energy-momenta and Einstein tensor components of
the non-linear electromagnetic model used.

\section{Colliding shock waves in Einstein-Maxwell (EM) theory}

The spacetime describing colliding em shock waves with general polarization
is summarized by \cite{5}%
\begin{multline}
ds^{2}=2e^{-M}dudv- \\
e^{-U}\left( \cosh W\left( e^{V}dx^{2}+e^{-V}dy^{2}\right) -2\sinh
Wdxdy\right) .
\end{multline}%
Here $\left( u,v\right) $ are the null coordinates while $M,$ $U,$ $V$ and $W
$ are metric functions depending on both $u$ and $v$ in the interaction
region. In the incoming regions, however, the metric functions depend only
on one (either $u$ or $v$) of the null coordinates. We must add that since
the waves are moving at the speed of light, their collision problem can be
best described in the null coordinates. The null coordinates are related to $%
\left( t,z\right) $ coordinates by $\sqrt{2}u=t+z$ and $\sqrt{2}v=t-z.$ The
incoming waves are moving along $\pm z$ and they collide at $t=z=0$ (or $%
u=v=0$). Since em waves are transverse in this picture, we expect to have
the $x$ and $y$ components of $\mathbf{E}$ (electric) and $\mathbf{B}$
(magnetic) vectors to be non-zero. The Maxwell and Einstein-Maxwell (EM)
equations must be satisfied with the appropriate boundary conditions. For
the EM waves these were formulated by O'Brien and Synge \cite{OS}, but in
the present problem, these conditions will be relieved. We wish also to
comment that the metric function $W$ carries the information about the
second (or cross, or relative) polarization of the incoming waves. Care
should be taken that a coordinate rotation of the $\left( x,y\right) $
coordinates must not yield parallelly polarized vectors in the two incoming
regions. Otherwise the waves are still linearly polarized so that the metric
function $W$ can be set to zero by a coordinate rotation. For a similar
situation in colliding gravitational waves, one may consult \cite{MH}.
Non-aligned polarization vectors in the incoming regions is therefore
crucial to obtain a genuine solution in the interaction region with $W\neq 0$%
. As a matter of fact, Bell and Szekeres gave the exact solution only with $%
W=0$ \cite{5}. With the exception of the case considered in Appendix C in
this paper, we shall restrict ourselves entirely to linear polarization. The
line element 
\begin{multline}
ds^{2}=2dudv-\cos ^{2}\alpha \left( u_{+}-v_{+}\right) dx^{2}- \\
\cos ^{2}\beta \left( u_{+}+v_{+}\right) dy^{2}
\end{multline}%
with the em potential $1-$form%
\begin{equation}
A\left( u,v\right) =\sin \alpha \left( u_{+}-v_{+}\right) dx+\sin \beta
\left( u_{+}+v_{+}\right) dy
\end{equation}%
solves the problem of colliding shock waves in EM theory subject to the
following information:

Here $\alpha $ and $\beta $ are amplitude constants of the em waves; $%
u_{+}=u\theta \left( u\right) $ and $v_{+}=v\theta \left( v\right) ,$ where $%
\theta \left( u\right) $ $/$ $\theta \left( v\right) $ is the Heaviside unit
step function. Note that we have the freedom to scale $u\rightarrow au$ and $%
v\rightarrow bv$ for constants $a$ and $b.$ Since we can absorb $ab$ into
the $x$ and $y$ coordinates and for the sake of simplicity, we shall make
the choice $a=b=1$ throughout the paper. For $u>0,v>0$, the line element (1)
represents the geometry of interaction region (region IV). The incoming
region II, for $v<0,$ $u>0$ is given by, (see Fig. 1) 
\begin{equation}
ds^{2}=2dudv-\cos ^{2}\left( \alpha u_{+}\right) dx^{2}-\cos ^{2}\left(
\beta u_{+}\right) dy^{2}.
\end{equation}%
For $u<0,$ $v>0$ we obtain the incoming region III from (1), which is
similar to II with $u\rightarrow v.$ The non-zero em field components are
obtained from (2) as follows;%
\begin{equation}
F_{ux}=\alpha \theta \left( u\right) \cos \alpha \left( u_{+}-v_{+}\right) ,
\end{equation}%
\begin{equation}
F_{vx}=-\alpha \theta \left( v\right) \cos \alpha \left( u_{+}-v_{+}\right) ,
\end{equation}%
\begin{equation}
F_{uy}=\beta \theta \left( u\right) \cos \beta \left( u_{+}+v_{+}\right) ,
\end{equation}%
\begin{equation}
F_{vy}=\beta \theta \left( v\right) \cos \beta \left( u_{+}+v_{+}\right) .
\end{equation}%
The Newman-Penrose (NP) quantities \cite{15} in the null basis $1-$forms 
\begin{equation}
\ell =du,\text{ \ \ \ }n=dv,
\end{equation}%
\begin{equation}
\sqrt{2}m=\cos \alpha \left( u_{+}-v_{+}\right) dx+i\cos \beta \left(
u_{+}+v_{+}\right) dy,
\end{equation}%
\begin{equation}
\sqrt{2}\bar{m}=\cos \alpha \left( u_{+}-v_{+}\right) dx-i\cos \beta \left(
u_{+}+v_{+}\right) dy
\end{equation}%
and their Ricci tensor connections are given in the Appendix A. From (5-8)
we can easily read the incoming em waves in region II (with $v<0$) and
region III (with $u<0$). Equivalently, we find the NP components of the em
field by 
\begin{multline}
\Phi _{2}\left( u\right) =F_{\mu \nu }\bar{m}^{\mu }n^{\nu }= \\
\frac{1}{\sqrt{2}}\left( \alpha +i\beta \right) \theta \left( u\right) \text{
\ \ (region II)}
\end{multline}%
\begin{multline}
\Phi _{0}\left( v\right) =F_{\mu \nu }\ell ^{\mu }m^{\nu }= \\
\frac{1}{\sqrt{2}}\left( \alpha +i\beta \right) \theta \left( v\right) \text{
\ \ \ (region III).}
\end{multline}%
The gravitational shock waves are also given in Appendix B as%
\begin{equation}
\Psi _{4}\left( u\right) =\frac{1}{2}\left( \alpha ^{2}-\beta ^{2}\right)
\theta \left( u\right) \text{\ \ (region II)}
\end{equation}%
\begin{equation}
\Psi _{0}\left( v\right) =\frac{1}{2}\left( \alpha ^{2}-\beta ^{2}\right)
\theta \left( v\right) \ \ \ \text{(region III).}
\end{equation}%
From (5-8), the electric and magnetic components of our fields are%
\begin{equation}
E_{x}=\frac{\alpha }{2}\left( \theta \left( u\right) -\theta \left( v\right)
\right) \cos \alpha \left( u_{+}-v_{+}\right) 
\end{equation}%
\begin{equation}
E_{y}=\frac{\beta }{2}\left( \theta \left( u\right) +\theta \left( v\right)
\right) \cos \beta \left( u_{+}+v_{+}\right) 
\end{equation}%
\begin{equation}
B_{x}=\frac{\beta }{2}\left( \theta \left( u\right) -\theta \left( v\right)
\right) \cos \beta \left( u_{+}+v_{+}\right) 
\end{equation}%
\begin{equation}
B_{y}=-\frac{\alpha }{2}\left( \theta \left( u\right) +\theta \left(
v\right) \right) \cos \alpha \left( u_{+}-v_{+}\right) .
\end{equation}

\begin{figure}[tbp]
\includegraphics[width=90mm,scale=0.7]{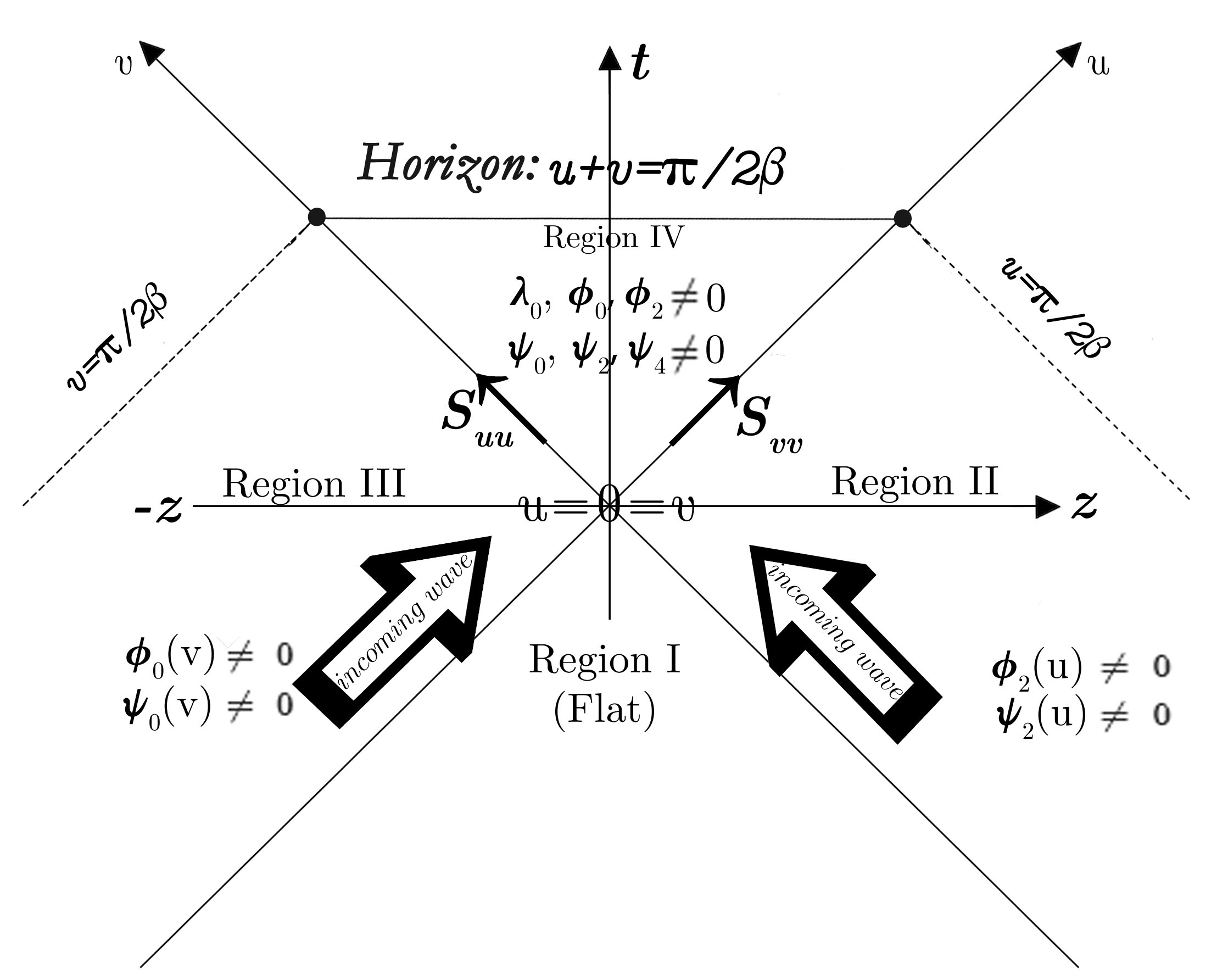} %
\captionsetup{justification=raggedright, singlelinecheck=false}
\caption{The spacetime diagram for colliding (em+grav) waves. The region I $%
(u<0,v<0)$ is flat, i.e. no-wave region. The incoming region II, $(u>0,v<0)$%
, has $\protect\phi _{2}\left( u\right) \neq 0\neq \protect\psi _{4}\left(
u\right) $ while region III $(u<0,v>0)$ has\ $\protect\phi _{0}\left(
v\right) \neq 0\neq \protect\psi _{0}\left( v\right) .$ The interaction
region IV $(u>0,v>0)$ has the non-zero components $\protect\phi _{0}\left(
u,v\right) ,\protect\phi _{2}\left( u,v\right) $ $\protect\psi _{0}\left(
u,v\right) ,\protect\psi _{4}\left( u,v\right) ,\protect\psi _{2}\left(
u,v\right) $ and $\protect\lambda _{0}$. The equivalent non-zero Ricci
tensor components are given in Appendix A. It can easily be seen that for ($%
\protect\alpha =\protect\beta $) in the waves sets $\protect\psi _{4}\left(
u\right) =0=\protect\psi _{0}\left( v\right) $ in the incoming regions and $%
\protect\lambda _{0}=0=\protect\psi _{2}$ in the interaction region. This
reduces the problem to colliding pure em waves problem of Bell and Szekeres.
We wish to draw attention in particular to the null sources $S_{uu}=\protect%
\delta \left( u\right) \left( \protect\beta \tan \protect\beta v_{+}-\protect%
\alpha \tan \protect\alpha v_{+}\right) $ and $S_{vv}=\protect\delta \left(
v\right) \left( \protect\beta \tan \protect\beta u_{+}-\protect\alpha \tan 
\protect\alpha u_{+}\right) $ emerging after the collision on the null
boundaries. These vanish for the choice $\protect\alpha =\protect\beta .$
From $S_{uu}$ and $S_{vv}$ we can read the null singularities as: $u=0,$ $%
v_{1}=\frac{\protect\pi }{2\protect\alpha },$ $v_{2}=\frac{\protect\pi }{2%
\protect\beta }$ and $v=0,$ $u_{1}=\frac{\protect\pi }{2\protect\alpha },$ $%
u_{2}=\frac{\protect\pi }{2\protect\beta }.$ }
\end{figure}
Since the incoming em waves move along $\pm z-$directions, they are
transverse. Therefore, their $E_{x},$ $E_{y}$ and $B_{x},$ $B_{y}$
components are all non-zero. This is due to the fact that the chosen ansatz
(3) for the vector potential is quite general. We note that our convention
to define the polarization of the em waves is based on the electric field
vector $\mathbf{E}$. An alternative choice can be made by using the magnetic
field vector $\mathbf{B}$. From our ansatz for the em potential (3) and the
derived field vectors (5-8), it is observed that the waves in region II and
III are linearly polarized. The $x-$components of $\mathbf{E}$ from region
II and III yield opposite signs, but this doesn't change the polarization.
We remind that em wave is a spin$-1$ field so that a $\pi -$rotation in
axis, i.e. from $+x$ to $-x$ is allowed. We finally note that $E_{x}\neq
E_{y}$ leads to elliptical polarization which reduces to circular
polarization for $E_{x}=E_{y}$, but these are still linearly polarized along
a line in the $xy-$plane. It is readily seen from (16-19) that after the
collision, we have $E_{x}=B_{x}=0.$ Prior to the collision, regions II and
III both have the em components elliptically polarized in the orthonormal
frame \{$u,v,\bar{x},\bar{y}$\}. To see this, we choose $d\bar{x}=g_{xx}dx$
and $d\bar{y}=g_{yy}dy$ so that 
\begin{equation}
\frac{E_{\bar{x}}^{2}}{\alpha ^{2}}+\frac{E_{\bar{y}}^{2}}{\beta ^{2}}=1.
\end{equation}%
Also note that, after the collision both the electric and magnetic fields
are polarized in $y-$direction, i.e. the process of collision acts as a
polarizer. Throughout the spacetime, EM field equations are given by 
\begin{equation}
R_{\mu \nu }-\frac{R}{2}g_{\mu \nu }=-T_{\mu \nu }+S_{\mu \nu }+\lambda
_{0}g_{\mu \nu }
\end{equation}%
where the energy-momentum tensor of the em field is 
\begin{equation}
T_{\mu \nu }=F_{\mu \lambda }F_{\nu }^{\;\lambda }-\frac{1}{4}g_{\mu \nu
}F_{\alpha \beta }F^{\alpha \beta }
\end{equation}%
and $S_{\mu \nu }$ stands for the energy-momentum on the null-hypersurfaces 
\cite{16}. From Appendix A and Eqs. (21-22) we read $S_{uu}=\delta \left(
u\right) \left( \beta \tan \beta v_{+}-\alpha \tan \alpha v_{+}\right) $ and 
$S_{vv}=\delta \left( v\right) \left( \beta \tan \beta u_{+}-\alpha \tan
\alpha u_{+}\right) $. The latter contains the delta functions of the Ricci
tensor as displayed in the Appendix. Let us add that upon suppressing the
infinite energy contributions from the planar $(x,y)$ directions the
integral of $S_{ab}$ can be shown to contribute finite to the $(u,v)$ plane.
This is a result of the integral 
\begin{multline}
2\int\nolimits_{-\delta }^{\delta }du\int_{0}^{\epsilon }dv\sqrt{-g}S_{uu}=
\\
\frac{\beta -\alpha }{\beta +\alpha }\left( 1-\cos \epsilon \left( \alpha
+\beta \right) \right) +\frac{\beta +\alpha }{\beta -\alpha }\left( 1-\cos
\epsilon \left( \beta -\alpha \right) \right) 
\end{multline}%
in which $\delta >0$ and $\epsilon >0$ are small parameters. A similar
result follows also from the $S_{vv}$ integral. The constant $\lambda _{0}$
is identified as the cosmological constant which turns out to be 
\begin{equation}
\lambda _{0}=\left( \alpha ^{2}-\beta ^{2}\right) \theta \left( u\right)
\theta \left( v\right) .
\end{equation}%
Obviously $\lambda _{0}$ emerges in region IV for $u>0$ and $v>0$ and
depending on whether $\alpha ^{2}>\beta ^{2}$ or $\alpha ^{2}<\beta ^{2}$ it
can be positive or negative. It is not difficult to speculate that the waves
may start with $\alpha ^{2}>\beta ^{2}$ but the $y-$mode can build up by
superposition or other mechanisms to suppress the $x$-mode in successive
collisions to make $\alpha ^{2}<\beta ^{2}$. Thus, the emergent cosmological
constant through colliding waves has the potential to change sign in
accordance with the dominance of linear $x$/$y$ modes.

From the Weyl scalars $\Psi _{4}\left( \Psi _{0}\right) $ (see Appendix B)
it can easily be seen that null singularities occur at $u=0$ ($v_{1}=\frac{%
\pi }{2\alpha }$ and $v_{2}=\frac{\pi }{2\beta }$) and $v=0$ ($u_{1}=\frac{%
\pi }{2\alpha }$ and $u_{2}=\frac{\pi }{2\beta }$) that is, they double in
number of the BS solution. When $\alpha =\beta $, the incoming Weyl
curvatures disappear and we recover the problem of colliding em shock waves
of BS \cite{5}.

The effect of second polarization on the formation of the cosmological
constant has also been considered in this study. The solution presented in 
\cite{7} is generalized to different amplitude wave profiles. Our analysis
has shown that the addition of second polarization does not yield a
cosmological constant. The related metric and Ricci scalar $\Lambda $ are
given in Appendix C.

\section{Colliding waves in Einstein-nonlinear electromagnetism}

Let's consider a general form of the nonlinear Maxwell Lagrangian as $%
\mathcal{L}\left( \mathcal{F}\right) $ in which $\mathcal{F}=F_{\mu \nu
}F^{\mu \nu }.$ Hence the Einstein nonlinear Maxwell action reads ($16\pi G=1
$) 
\begin{equation}
S=\int d^{4}x\sqrt{-g}\left[ \frac{R}{2}+\mathcal{L}\left( \mathcal{F}%
\right) \right] .
\end{equation}%
The line element is chosen to be (note that in this section we use more
appropriately the commonly used $+2$ signature) 
\begin{equation}
ds^{2}=-2dudv+\cos ^{2}\xi dx^{2}+dy^{2}
\end{equation}%
in which%
\begin{equation}
\xi =\epsilon \left( u_{+}-v_{+}\right) .
\end{equation}%
Our em potential ansatz is $\mathbf{A}=a_{0}\sin \xi dx$ so that the em
field $2-$form is%
\begin{equation}
\mathbf{F}=\alpha \cos \xi \left( \mathcal{\theta }\left( u\right) du-%
\mathcal{\theta }\left( v\right) dv\right) \wedge dx,
\end{equation}%
where the constant $\alpha $ is $\alpha =a_{0}\epsilon .$ Its dual $^{\ast }%
\mathbf{F}$ is given by 
\begin{equation}
^{\ast }\mathbf{F=}-\alpha \left( \mathcal{\theta }\left( u\right) du+%
\mathcal{\theta }\left( v\right) dv\right) \wedge dy
\end{equation}%
and 
\begin{equation}
\mathcal{F}=4\alpha ^{2}\mathcal{\theta }\left( u\right) \mathcal{\theta }%
\left( v\right) .
\end{equation}%
The final form of $\mathcal{F}$ implies that $\mathcal{F}$ is non-zero only
in the interaction region i.e. $v>0$ and $u>0$ (i.e. the incoming em fields
are null) and it is a constant. We must add that choosing the field ansatz
as in (28) guarantees that the other Maxwell invariant is zero i.e., $%
\mathcal{G}=-\frac{1}{4}F_{\mu \nu }$ $^{\ast }F^{\mu \nu }=0.$ Therefore
the general form of the nonlinear Lagrangian depends only on $\mathcal{F}$.
For instance, in the case of BI theory, Lagrangian becomes 
\begin{equation}
\mathcal{L}\left( \mathcal{F}\right) =2b^{2}\left( 1-\sqrt{1+\frac{\mathcal{F%
}}{b^{2}}}\right) 
\end{equation}%
which reduces to the linear Maxwell Lagrangian in the limit $b\rightarrow
\infty $. We note that $b\neq 0$ is called the BI parameter with dimension
of mass. The nonlinear Maxwell's equation in the interaction region ($u>0$
and $v>0$) is given by%
\begin{equation}
d\left( ^{\ast }\mathbf{F}\frac{d\mathcal{L}\left( \mathcal{F}\right) }{d%
\mathcal{F}}\right) =0.
\end{equation}%
or effectively for (29) and (31) it means 
\begin{equation}
d\left( ^{\ast }\mathbf{F}\right) =0,
\end{equation}%
which is obviously satisfied. On the boundaries, however, this gives null
currents i.e. 
\begin{equation}
d\left( ^{\ast }\mathbf{F}\frac{d\mathcal{L}\left( \mathcal{F}\right) }{d%
\mathcal{F}}\right) =\text{ }^{\ast }\mathbf{J}
\end{equation}%
where $^{\ast }\mathbf{J}$ is the current $3-$form given by%
\begin{equation}
^{\ast }\mathbf{J=}\frac{2\alpha ^{3}}{b^{2}}\frac{\left( \delta \left(
u\right) \theta \left( v\right) -\delta \left( v\right) \theta \left(
u\right) \right) }{\left( 1+\frac{4\alpha ^{2}}{b^{2}}\theta \left( u\right)
\theta \left( v\right) \right) ^{3/2}}du\wedge dv\wedge dy,
\end{equation}%
which occurs similar to $S_{ab}$ on the null boundaries. The energy momentum
tensor of this nonlinear field and its explicit components are given in
Appendix D. Plugging these into the field equations inside the interaction
region 
\begin{equation}
G_{\mu }^{\nu }-\lambda _{0}\delta _{\mu }^{\nu }=T_{\mu }^{\nu }+S_{\mu
}^{\nu }
\end{equation}%
yields $\lambda _{0}=-\left( \mathcal{L}-2\mathcal{F}\frac{d\mathcal{L}}{d%
\mathcal{F}}\right) $ and to have consistency with the other equations, we
must have $\mathcal{\epsilon }^{2}=-4\alpha ^{2}\frac{d\mathcal{L}\left( 
\mathcal{F}\right) }{d\mathcal{F}}$ which is a constant (note that $\frac{d%
\mathcal{L}\left( \mathcal{F}\right) }{d\mathcal{F}}<0$). The fact that $%
\lambda _{0}$ emerges as a constant is by virtue of the chosen Lagrangian
(31) and the solution (26-28).

Breton considered the following line element \cite{10} 
\begin{equation}
ds^{2}=-2dudv+\cos ^{2}\xi dx^{2}+\cos ^{2}\eta dy^{2}
\end{equation}%
in which 
\begin{equation}
\xi =\epsilon \left( u_{+}+v_{+}\right) 
\end{equation}%
and 
\begin{equation}
\eta =\kappa \left( u_{+}-v_{+}\right) 
\end{equation}%
together with the em $2-$form (here $\epsilon $ and $\kappa $ are amplitude
constants analogous to our $\alpha $ and $\beta $ in the linear Maxwell
theory)%
\begin{multline}
\mathbf{F=}\alpha _{0}\cos \xi \left( dx\wedge du+dx\wedge dv\right) - \\
\beta _{0}\cos \eta \left( dy\wedge du-dy\wedge dv\right) .
\end{multline}%
Note that the constants $\alpha _{0}$ and $\beta _{0}$ are related to $%
\epsilon $ and $\kappa $ through the field equations. The nonlinear
Lagrangian used here is the more general BI Lagrangian given by%
\begin{equation}
\mathcal{L}=2b^{2}\left( 1-\sqrt{1+\frac{\mathcal{F}}{b^{2}}-\frac{\mathcal{G%
}^{2}}{b^{4}}}\right) .
\end{equation}%
As it was shown in Ref. \cite{10} in the limit $b\rightarrow \infty $ it
reduces rightly to the Bell-Szekeres solution. In this general case we also
find that there is emergent cosmological constant in the interaction region.
Emergence of null currents on the null boundaries after collision, however,
from the nonlinear Maxwell equation is also inevitable in \cite{10}. For the
case of different nonlinear electromagnetic models other than BI, a similar
conclusion remains to be seen.

\section{Conclusion}

In this study we propose that cosmological constant is of em origin.
Collision of linearly polarized em waves accompanied by appropriate
gravitational shock waves gives rise to cosmological constant $\lambda
_{0}=\alpha ^{2}-\beta ^{2}$ in the interaction region ($u>0$ and $v>0$).
Here $\alpha $ and $\beta $ are amplitude constants of the incoming em
waves. For $\alpha =\beta $ we have the typical collision of em shock waves
derived first by Bell and Szekeres \cite{5} in which the interaction region
has only em field with null singularities on the null boundaries. Two null
em waves collide and turn into a non-null em field which is isometric to the
Bertotti-Robinson spacetime (see \cite{1}). Now, an interesting situation,
observed by Barrabes and Hogan \cite{4} arises: When the wave amplitudes
along two space directions are unequal (i.e. $\alpha \neq \beta $), a
cosmological constant emerges in the interaction region. We use this
observation to speculate about the possible origin of the cosmological
constant. We prove that this happens when the em waves are linearly
polarized. We do this by extending the problem to cross polarized collision
where the em energy transforms into a general form of energy-momentum which
can't be identified as a cosmological constant. (See Appendix C)

We also show that the collision of em waves in a nonlinear electromagnetism,
specifically a reduced version of the BI theory, similar trace of
cosmological constant emerges. It should be added that the nonlinear Maxwell
equations are satisfied modulo the currents on the null boundaries, after
the collision, much like the sources $S_{ab}$ of the linear theory on the
null-boundaries.

\textbf{Appendix A:}

From the metric (1) and NP null-tetrad (8-10) we obtain the following Ricci
components 
\begin{multline}
2\phi _{22}=R_{u\,u}=\left( \alpha ^{2}+\beta ^{2}\right) \theta \left(
u\right) + \\
\delta \left( u\right) \left( \beta \tan \left( \beta v_{+}\right) -\alpha
\tan \left( \alpha v_{+}\right) \right)
\end{multline}%
\begin{multline}
2\phi _{00}=R_{vv}=\left( \alpha ^{2}+\beta ^{2}\right) \theta \left(
v\right) + \\
\delta \left( v\right) \left( \beta \tan \left( \beta u_{+}\right) -\alpha
\tan \left( \alpha u_{+}\right) \right)
\end{multline}%
\begin{equation}
2\phi _{02}=R_{\mu \nu }m^{\mu }m^{\nu }=\left( \alpha ^{2}+\beta
^{2}\right) \theta \left( v\right) \theta \left( v\right)
\end{equation}%
\begin{equation}
R_{u\,v}=\left( \beta ^{2}-\alpha ^{2}\right) \theta \left( u\right) \theta
\left( v\right)
\end{equation}%
\begin{equation}
R_{xx}=2\alpha ^{2}\theta \left( u\right) \theta \left( v\right) \cos
^{2}\alpha \left( u-v\right)
\end{equation}%
\begin{equation}
R_{yy}=-2\beta ^{2}\theta \left( u\right) \theta \left( v\right) \cos
^{2}\beta \left( u+v\right)
\end{equation}%
\begin{equation}
R=-24\Lambda =4\left( \beta ^{2}-\alpha ^{2}\right) \theta \left( u\right)
\theta \left( v\right) =-4\lambda _{0}
\end{equation}%
\begin{eqnarray*}
&&\text{(}\delta \left( u\right) \text{ and }\delta \left( v\right) \text{
are Dirac delta functions and } \\
u_{+} &=&u\theta \left( u\right) \text{ and }v_{+}=v\theta \left( v\right) 
\text{)}
\end{eqnarray*}

\textbf{Appendix B:}

The non-zero Weyl components $\psi _{2},\psi _{4}$ and $\psi _{0}$ are as
follows:%
\begin{equation}
6\psi _{2}=\left( \alpha ^{2}-\beta ^{2}\right) \theta \left( u\right)
\theta \left( v\right)
\end{equation}%
\begin{multline}
2\psi _{4}=\left( \alpha ^{2}-\beta ^{2}\right) \theta \left( u\right) - \\
\delta \left( u\right) \left( \alpha \tan \left( \alpha v_{+}\right) +\beta
\tan \left( \beta v_{+}\right) \right)
\end{multline}%
\begin{multline}
2\psi _{0}=\left( \alpha ^{2}-\beta ^{2}\right) \theta \left( v\right) - \\
\delta \left( v\right) \left( \alpha \tan \left( \alpha u_{+}\right) +\beta
\tan \left( \beta u_{+}\right) \right)
\end{multline}%
\textbf{Appendix C:}

\textit{Colliding em waves with cross polarization:}

Collision of linearly polarized em waves was generalized to include the
second polarization in \cite{6,7}. The situation is analogous to
Khan-Penrose \cite{8} and Nutku-Halil \cite{9}, or Schwarzschild-Kerr
relation. The latter solutions contain one extra parameter so that when the
parameter vanishes, we obtain the former solutions. In the wave collision
problem, the parameter is the angle of relative polarization of the two
incoming waves, say $\alpha _{0}$. The metric of the interaction region
induces a cross-term $g_{xy}$ which is proportional to $\sin \alpha _{0}$ so
that $g_{xy}\rightarrow 0$, when the waves are linearly polarized. With
reference to \cite{6}, it is not difficult to give the spacetime of the
interaction region in oblate-spheroidal type coordinates \cite{9}%
\begin{multline}
ds^{2}=F\left( \frac{d\tau ^{2}}{\beta ^{2}\Delta }-\frac{d\sigma ^{2}}{%
\alpha ^{2}\delta }\right) -\frac{\delta }{F}dx^{2}- \\
\left( \Delta F+\frac{\delta }{F}\left( \tau \sin \alpha _{0}\right)
^{2}\right) dy^{2}+\frac{2\tau \delta }{F}\sin \alpha _{0}dxdy.
\end{multline}%
The notation here goes as follows%
\begin{equation}
\tau =\sin \beta \left( u+v\right) ,\text{ \ \ }\sigma =\sin \alpha \left(
u-v\right) ,
\end{equation}%
\begin{equation}
\Delta =1-\tau ^{2},\text{ \ \ }\delta =1-\sigma ^{2},
\end{equation}%
\begin{equation}
2F=\sqrt{1+\sin ^{2}\alpha _{0}}\left( 1+\sigma ^{2}\right) +1-\sigma ^{2}.
\end{equation}%
Note that 
\begin{equation}
\frac{d\tau ^{2}}{\beta ^{2}\Delta }-\frac{d\sigma ^{2}}{\alpha ^{2}\delta }%
=4dudv
\end{equation}%
and in the limit $\alpha _{0}\rightarrow 0$ we recover the BS metric of
linear polarization. The existence of $\alpha \neq \beta ,$ however, makes
the curvature components rather complicated so that a pure cosmological
constant doesn't arise in the present case. To verify this we make use of
the null-tetrad basis 1-forms 
\begin{equation}
\sqrt{2}\ell =\sqrt{F}\left( \frac{d\tau }{\beta \sqrt{\Delta }}-\frac{%
d\sigma }{\alpha \sqrt{\delta }}\right) 
\end{equation}%
\begin{equation}
\sqrt{2}n=\sqrt{F}\left( \frac{d\tau }{\beta \sqrt{\Delta }}+\frac{d\sigma }{%
\alpha \sqrt{\delta }}\right) 
\end{equation}%
\begin{equation}
\sqrt{2}m=\sqrt{\frac{\delta }{F}}\left( dx-\tau \sin \alpha _{0}dy\right) +i%
\sqrt{\Delta F}dy
\end{equation}%
and the complex conjugate of $m$. It suffices to compute the NP scalar $%
\Lambda $ which is 
\begin{equation}
\Lambda =\frac{\left( \alpha ^{2}-\beta ^{2}\right) }{48F^{3}}\left(
4F^{2}-\delta \sin ^{2}\alpha _{0}\right) .
\end{equation}%
From the trace of equation (20), we see that the expected cosmological
'constant' $\lambda _{0}=-\frac{R}{4}$ is not a constant. In the linear
polarization limit $\alpha _{0}=0$ ($F=1$) we obtain that $\lambda _{0}=$ $%
\frac{1}{2}\left( \alpha ^{2}-\beta ^{2}\right) .$ (Note that the extra $%
\frac{1}{2}$ factor comes from $4dudv$ in (3) instead of $2dudv$). Therefore
we conclude that emergence of cosmological constant is related to the linear
polarization property of the colliding waves. We recall that the cross
polarization of waves through the Faraday rotation may be instrumental in
the detection of gravitational waves \cite{3}.

\textbf{Appendix D:}

The energy-momentum tensor $T_{\mu }^{\nu }$ in region IV with its explicit
components and Einstein's terms in all regions are given as follow:

\begin{equation}
T_{\mu }^{\nu }=\mathcal{L}\delta _{\mu }^{\nu }-4F_{\mu \lambda }F^{\nu
\lambda }\frac{d\mathcal{L}}{d\mathcal{F}}
\end{equation}%
\begin{equation}
T_{u}^{u}=\mathcal{L}-\mathcal{F}\frac{d\mathcal{L}}{d\mathcal{F}},
\end{equation}%
\begin{equation}
T_{v}^{v}=\mathcal{L}-\mathcal{F}\frac{d\mathcal{L}}{d\mathcal{F}},
\end{equation}%
\begin{equation}
T_{u}^{v}=4\alpha ^{2}\theta \left( u\right) \frac{d\mathcal{L}}{d\mathcal{F}%
},\text{ \ \ }T_{v}^{u}=4\alpha ^{2}\theta \left( v\right) \frac{d\mathcal{L}%
}{d\mathcal{F}},
\end{equation}%
\begin{equation}
T_{x}^{x}=\mathcal{L}-2\mathcal{F}\frac{d\mathcal{L}}{d\mathcal{F}},\text{ \
\ }T_{y}^{y}=\mathcal{L}.
\end{equation}%
\begin{equation}
G_{u}^{u}=-\epsilon ^{2}\theta \left( u\right) \theta \left( v\right) ,\text{
\ \ }G_{v}^{v}=-\epsilon ^{2}\theta \left( u\right) \theta \left( v\right) ,
\end{equation}%
\begin{equation}
G_{u}^{v}=-\epsilon ^{2}\theta \left( u\right) -\epsilon \tan \xi \delta
\left( u\right) ,\text{ \ }
\end{equation}%
\begin{equation}
G_{v}^{u}=-\epsilon ^{2}\theta \left( v\right) +\epsilon \tan \xi \delta
\left( v\right) ,
\end{equation}%
\begin{equation}
G_{x}^{x}=0,\text{ \ }G_{y}^{y}=-2\epsilon ^{2}\theta \left( u\right) \theta
\left( v\right) .
\end{equation}

\end{document}